**September 06, 2025**

## Cross-sectoral emission interdependencies of waste management

Author 1

- Nikolaos Kalyviotis Dipl.-Ing., MSc, MSPM, MBA, PhD
- Assistant Professor of Planning, Management and Evaluation of Programs, Investments and Projects, Department of Planning and Regional Development, School of Engineering, University of Thessaly, Volos, Greece
- https://orcid.org/0009-0006-5658-4262

Author 2

- Georgina Calypso Kalogerakis Dipl.-Ing., PhD
- Postdoctoral Fellow, Department of Chemical Engineering, Faculty of Engineering, McGill University, Montreal, Canada
- https://orcid.org/0000-0002-9438-9525

Author 3

- Georgios Kolliopoulos Dipl.-Ing., PhD
- Associate Professor, Department of Mineral, Metallurgical and Materials Engineering, Faculty of Science and Engineering, Université Laval, Quebec, Canada
- https://orcid.org/0000-0003-3653-3362

**CORRESPONDENCE Nikolaos Kalyviotis nkalyviotis@uth.gr**

**Abstract**

This study examines emission interdependencies within the waste sector and between waste and other economic domains—energy, transport, water, and communications—using Life Cycle Assessment (LCA) methodologies with emphasis on process-based and hybrid approaches. The analysis employs the EXIOBASE multi-regional input-output database, encompassing 1,260 processes across five key sectors. Statistical correlation is applied to identify commonalities and divergences in emission drivers. Results indicate strong correlations among certain landfill waste streams, particularly organic-based materials, as well as links between plastics and wood, suggesting shared decomposition or management pathways. Moderate correlations appear in cross-sectoral interactions, especially between waste, energy, and transport, while weak correlations are observed in processes with distinct treatment mechanisms, such as landfill versus wastewater. Methodologically, the study highlights challenges posed by high-dimensional datasets, including multicollinearity, which limit the effectiveness of linear models. These findings underscore the complexity of sectoral interdependencies and the need for both integrated and material-specific strategies. By demonstrating how systemic tools like EXIOBASE can capture these linkages, the research provides a foundation for constructing more robust emission models and informing targeted mitigation policies. The results contribute to advancing environmental accounting frameworks while offering practical insights for emission reduction planning across interconnected sectors.

**Keywords**
UN SDG 6: Clean Water and Sanitation; UN SDG 12: Responsible Consumption and Production; UN SDG 13: Climate Action; UN SDG 14: Life Below Water; Waste Management and Disposal; Waste

## 1. Introduction

The global challenge of mitigating emissions necessitates understanding interdependencies across economic sectors (Kalyviotis et al., 2024). Waste management, including landfill, incineration, and wastewater treatment, is a significant CO2 source due to organic decomposition and energy-intensive processes (Barati et al. 2022). Energy production, transport, and water supply sectors also contribute substantially to emissions, often interlinked through supply chains (Pan, 2015). This study leverages the EXIOBASE 3 database (Stadler et al., 2018), a multi-regional environmentally extended input-output (MR EE IO) framework, to analyse CO2 emission correlations across waste management and other sectors, including energy (e.g., coal, nuclear), transport (e.g., rail, air), water supply, and telecommunications.

Previous studies identified strong relationships among certain landfill waste categories, moderate relationships across others, and comparatively weak relationships between landfill and wastewater treatment (see Ragab et al., 2025; Ingle et al., 2025; Madhavaraj & Karthikeyan, 2025; Merab et al., 2025; Bhuiya et al., 2025; Bugarčić et al., 2024; Diamond, 2024; Santos et al., 2024; Riman et al., 2022; Singh & Chunglok, 2022; Aziz et al., 2020; Meereboer et al., 2020; Dos Santos et al., 2020; Souza et al., 2020; Abouri et al., 2016; Bohlmann, 2005; Owens & Chynoweth, 1993). New data expands these analyses, reporting correlation coefficients across 163 Input-Output Groups (IOGs). This paper synthesizes these findings to explore emission dependencies, assess their alignment with academic literature, and propose policy implications.

The joint effort between the three authors (see Kalyviotis, Kalogerakis & Kolliopoulos, 2025) emerged from prior research conducted during the doctoral studies of the first author, which focused on the reduction of emissions in the transport sector (see Kalyviotis, 2022). Specifically, that work investigated which transport-related economic activities should be prioritized in order to maximize emission reduction potential (Kalyviotis et al., 2018). For this purpose, the EXIOBASE 3 database was employed to examine the interdependencies of the transport sector with the water, energy, waste, and communication sectors (Kalyviotis Rogers & Hewings, 2025). The analysis demonstrated that the energy and waste sectors present comparatively greater potential for further emission reductions. Building on this insight, the research was subsequently extended to focus on the waste sector. While the first author possessed expertise in energy and transport related modelling, there was a recognized need for complementary expertise in the waste sector to ensure a robust interpretation of the results. To address this gap, the collaboration with the other two authors was initiated. Both collaborators had been previously known to the first author through academic tenures at the University of Toronto, and their specialised knowledge in the waste sector provided the necessary foundation for the present work.

## 2. Methodology overview and data framework

The classification of Life Cycle Assessment (LCA) methodologies can be conceptualized along a continuum that balances data intensity, and the level of uncertainty or assumptions involved. As illustrated in the figure, five primary LCA approaches are identified: Process-based, Hybrid, Pseudo, Simplified, and Parametric (Figure 1).

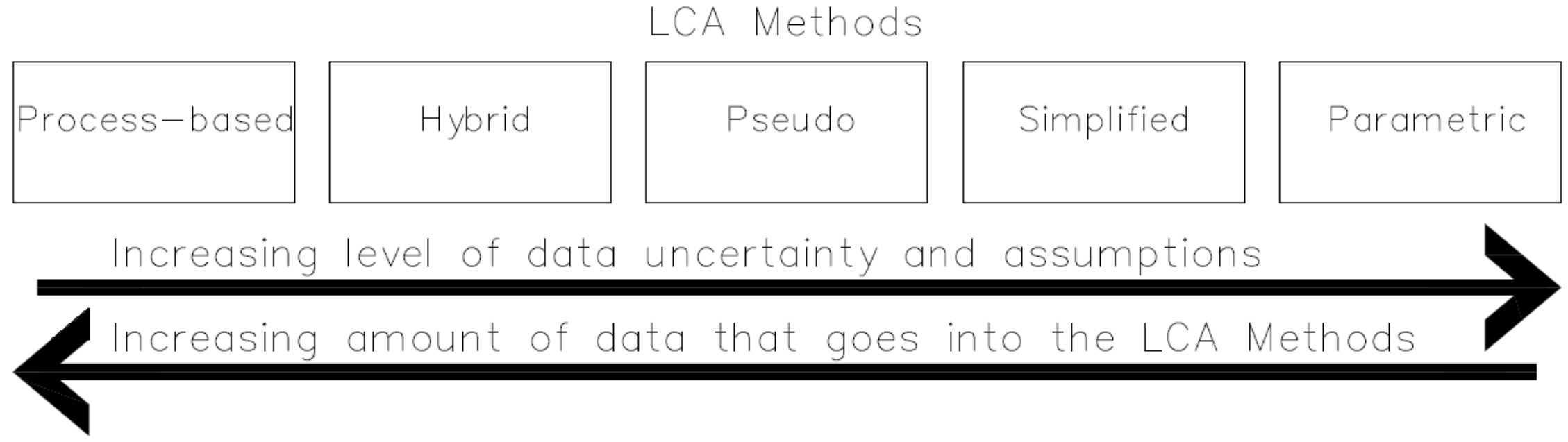

Figure 1 LCA Methods Ranking based on Data Requirements and Uncertainty (Kalyviotis, Rogers & Hewings, 2025)

At one end of the spectrum, Process-based LCA requires a substantial volume of detailed data and is typically characterized by lower uncertainty and fewer assumptions (Pomponi & Lenzen, 2018; Yang et al., 2017). Process-based LCA is the most traditional and detailed form of life cycle assessment (Bamber et al., 2020). This method provides high-resolution environmental assessments by modeling individual processes throughout a product's life cycle (Li et al., 2025; Baral & Bakshi, 2006). It models each step in the life cycle of a product or service using specific input and output data for individual processes (e.g., material extraction, manufacturing, transport, use, disposal) (Schaubroeck & Benetto, 2024). This method provides high-resolution environmental results but is typically data-intensive, time-consuming, and often subject to system boundary truncation (Shi et al., 2024; Stephan et al., 2019; Plepys & Sing, 2019). Moving along the continuum, Hybrid LCA integrates process-based data with economic input-output data, thereby extending system boundaries while managing data demands (Jakobs, 2023; Ghosh & Bakshi, 2020; Pomponi & Lenzen, 2018; Suh et al., 2004). Pseudo LCA further simplifies data requirements and tends to involve more assumptions, trading off some accuracy for broader applicability (Olugbenga et al., 2019). Simplified LCA and Parametric LCA represent approaches that significantly reduce data inputs, often relying on generalized parameters or streamlined models (Eleftheriou et al., 2022; Kiemel et al., 2022; Hollberg & Ruth, 2016). These methods are associated with higher levels of uncertainty and are typically employed for preliminary assessments or comparative studies where comprehensive data is unavailable (Jolivet et al., 2021; Hollberg & Ruth 2016). Figure 1 underscores the inverse relationship between data input and uncertainty: as the amount of data incorporated into the LCA decreases, the reliance on assumptions and the potential for uncertainty increases. This trade-off is central to selecting the most appropriate LCA method based on the objectives of the study, available data, and required precision.

A comparative analysis between process-based Life Cycle Assessment (LCA) and the EXIOBASE database reveals key differences in methodology, data scope, and usability. Process-based LCA is characterized by its relatively straightforward simple methodology (Ayres, 1995). Although it is often time- and resource-intensive, it provides a high level of environmental detail for individual products (Beylot et al., 2020). However, its scope is typically limited to specific areas of consumption and material flows (Castellani et al., 2019). In terms of data sources, process-based LCA relies on a mix of open databases and private industrial data, which may sometimes be confidential (Finkbeiner et al., 2006). Furthermore, it generally lacks integrated socio-economic data (Gutowski, 2018).

In contrast, EXIOBASE offers a high level of sectoral disaggregation (Stadler et al., 2018) and supports both micro- and macro-scale environmental accounting (Merciai & Schmidt, 2017). The database encompasses a broad range of products and consumption areas within a unified framework (Stadler et al., 2018). EXIOBASE includes monetary Input-Output Tables (MR-IOT) and Multi-Regional Environmentally Extended Supply-Use Tables (MR-SUT), facilitating detailed supply-use analyses (Beylot et al., 2020; Tukker et al., 2018). Moreover, it integrates process-based LCA data and provides coefficients and statistical outputs, enabling comprehensive environmental assessments (Beylot et al., 2020; Tukker et al., 2018).

This comparison highlights the complementary nature of both approaches: while process-based LCA excels in granularity at the product level, EXIOBASE offers extensive coverage and systemic insights into environmental impacts across economies, which this research also aims to provide.

EXIOBASE 3 is a comprehensive, global Multi-Regional Environmentally Extended Input-Output (MR EE IO) database that encompasses 44 countries and 5 Rest-of-the-World (RoW) regions (Stadler et al., 2018). It provides detailed data on 163 industries and 200 product categories for the period 1995–2011, with nowcasted extensions available through 2022 (Owen et al., 2025; Rasul et al., 2024). The database includes information on emissions and resource use (e.g., $CO_2$ emissions, material extraction) by industry, integrated via international trade flows to facilitate the assessment of environmental footprints (Wood et al., 2018;).

EXIOBASE 3 is particularly well-suited for analyzing emissions associated with various waste categories—such as food, textiles, paper, wood, plastics, and wastewater—due to its ability to trace material flows and environmental impacts throughout global supply chains (Leclerc et al., 2023; Zeller et al., 2018).

The raw dataset includes supply-use tables (SUTs) and emissions accounts (Stadler et al., 2018). This data was used to calculate correlations between specific waste streams and associated emissions. The methodology adopted in this study follows an inductive (see Figure 2) rather than a deductive approach (see D'Incognito et al., 2015), a choice motivated by the extensive number of processes under consideration. In total, the analysis encompassed five economic sectors, corresponding to 1,260 emission factors and thus 1,260 distinct processes.

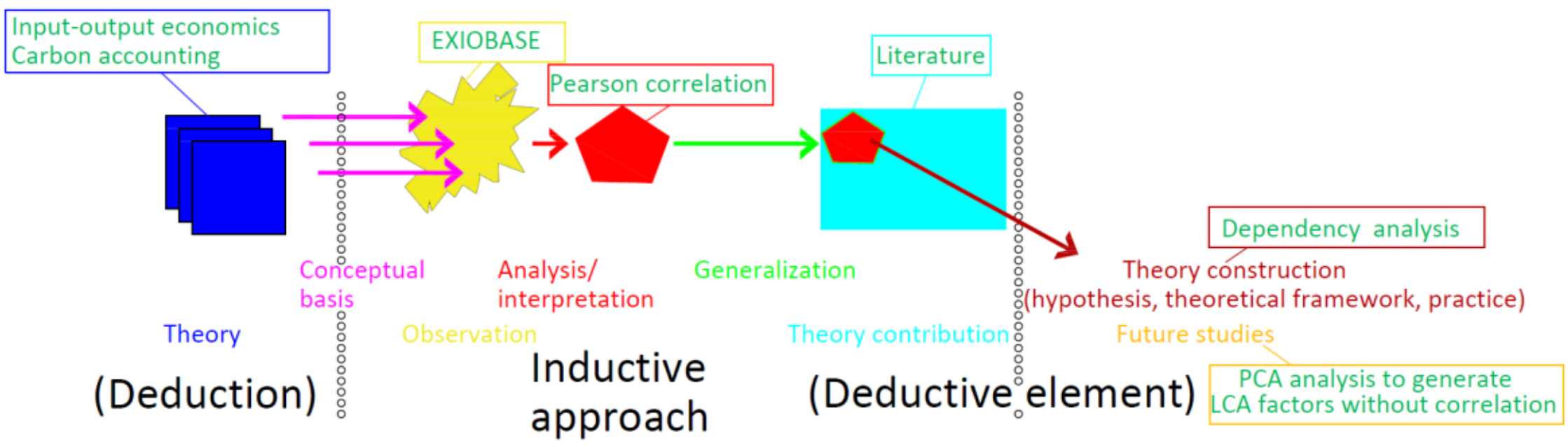

Figure 2 Inductive research workflow adopted in this research

The first step of the analysis relied on the EXIOBASE 3 database, the theoretical framework of which was accepted as the foundation for this research, given its development through the collaboration of multiple partner organizations and a broad network of researchers and experts. Building on this established framework, the dataset was initially examined using Pearson correlation (Field, 2024) in order to identify patterns of interdependence among sectors. The present analysis focuses on statistical correlations—quantified using p-values (although it is more conventional to report correlation coefficients with p-values indicating statistical significance)—between emissions from landfill and wastewater treatment across different waste categories. The detailed sectoral and product-level structure of EXIOBASE 3 enables such analysis by providing disaggregated data on emissions from waste management sectors (e.g., landfills, wastewater treatment) as well as from material-specific industries (e.g., food production, textile manufacturing). The results were then re-examined in light of EXIOBASE's theoretical assumptions, through the theoretical framework presented in Chapter 4, enabling refinement of the analytical framework and exploration of potential generalizations.

To establish a comprehensive dataset, emission data were extracted from five key economic sectors—waste, transport, communication, water, and energy (see Table 1). Within this framework, waste categories were systematically mapped to EXIOBASE 3 product classifications to ensure consistency and comparability across sectors.

Table 1 Emissions generation correlation between sectors

| Input-Output Groups' dependency/connection** | Correlation | Sample size* | Type of dependency |
|---|---|---|---|
| Incineration of waste: Plastic - Incineration of waste: Textiles | 1.000 | x | Waste sector interdependency |
| Composting of food waste, incl. land application - Biogasification of paper, incl. land application | 0.999 | x | Waste sector interdependency |
| Incineration of waste: Metals and Inert Materials - Composting of food waste, incl. land application | 0.999 | x | Waste sector interdependency |
| Incineration of waste: Metals and Inert Materials - Incineration of waste: Oil/Hazardous waste | 0.993 | x | Waste sector interdependency |
| Steam and hot water supply - Incineration of waste: Metals and Inert Materials | 0.993 | x | Water-Waste sectors dependency |
| Steam and hot water supply - Composting of food waste, incl. land application | 0.992 | x | Water-Waste sectors dependency |
| Landfill of waste: Food - Landfill of waste: Textiles | 0.991 | ✓ | Waste sector interdependency |
| Landfill of waste: Paper - Landfill of waste: Textiles | 0.990 | ✓ | Waste sector interdependency |
| Landfill of waste: Food - Landfill of waste: Paper | 0.990 | ✓ | Waste sector interdependency |
| Landfill of waste: Plastic - Landfill of waste: Wood | 0.957 | ✓ | Waste sector interdependency |
| Incineration of waste: Oil/Hazardous waste - Composting of food waste, incl. land application | 0.928 | x | Waste sector interdependency |
| Steam and hot water supply - Landfill of waste: Inert / Metal / Hazardous | 0.919 | ✓ | Water-Waste sectors dependency |
| Steam and hot water supply - Incineration of waste: Oil/Hazardous waste | 0.919 | x | Water-Waste sectors dependency |
| Incineration of waste: Metals and Inert Materials - Landfill of waste: Inert / Metal / Hazardous | 0.919 | x | Waste sector interdependency |
| Transmission of electricity - Distribution and trade of electricity | 0.918 | ✓ | Energy sector interdependency |
| Composting of food waste, incl. land application - Landfill of waste: Inert / Metal / Hazardous | 0.913 | x | Waste sector interdependency |
| Landfill of waste: Food - Landfill of waste: Wood | 0.893 | ✓ | Waste sector interdependency |
| Landfill of waste: Paper - Landfill of waste: Wood | 0.888 | x | Waste sector interdependency |
| Incineration of waste: Oil/Hazardous waste - Landfill of waste: Inert / Metal / Hazardous | 0.887 | x | Waste sector interdependency |
| Incineration of waste: Food - Incineration of waste: Paper | 0.874 | x | Waste sector interdependency |
| Landfill of waste: Textiles - Landfill of waste: Wood | 0.872 | ✓ | Waste sector interdependency |
| Production of electricity by coal - Distribution and trade of electricity | 0.846 | ✓ | Energy sector interdependency |
| Transmission of electricity - Collection, purification and distribution of water | 0.841 | ✓ | Energy-Water sectors dependency |
| Production of electricity by coal - Transmission of electricity | 0.828 | ✓ | Energy sector interdependency |

| Wastewater treatment, food - Wastewater treatment, other | 0.816 | ✓ | Waste sector interdependency |
|---|---|---|---|
| Manufacture of motor vehicles, trailers and semi-trailers - Transport via pipelines | 0.803 | ✓ | Transport sector interdependency |
| Production of electricity by nuclear - Production of electricity by wind | 0.767 | *x* | Energy sector interdependency |
| Distribution and trade of electricity - Collection, purification and distribution of water | 0.761 | ✓ | Energy-Water sectors dependency |
| Landfill of waste: Paper - Landfill of waste: Plastic | 0.753 | ✓ | Waste sector interdependency |
| Landfill of waste: Food - Landfill of waste: Plastic | 0.750 | ✓ | Waste sector interdependency |
| Production of electricity by tide, wave, ocean - Composting of paper and wood, incl. land application | 0.731 | *x* | Energy-Waste sectors dependency |
| Sale, maintenance, repair of motor vehicles, motor vehicles parts, motorcycles, motorcycles parts and accessories - Post and telecommunications | 0.725 | ✓ | Transport-Communication sectors dependency |
| Landfill of waste: Plastic - Landfill of waste: Textiles | 0.715 | ✓ | Waste sector interdependency |
| Production of electricity by coal - Collection, purification and distribution of water | 0.707 | ✓ | Energy-Water sectors dependency |
| Transport via railways – Post and telecommunications | 0.617 | ✓ | Transport-Communication sectors dependency |
| Production of electricity by petroleum and other oil derivatives - Air transport | 0.598 | ✓ | Transport-Energy sectors dependency |
| Wastewater treatment, food - Landfill of waste: Food | 0.591 | ✓ | Waste sector interdependency |
| Sea and coastal water transport - Incineration of waste: Plastic | 0.591 | *x* | Transport-Waste sectors dependency |
| Sea and coastal water transport - Incineration of waste: Textiles | 0.591 | *x* | Transport-Waste sectors dependency |
| Production of electricity by geothermal - Transport via pipelines | 0.589 | *x* | Energy-Transport sectors dependency |
| Wastewater treatment, food - Landfill of waste: Paper | 0.579 | ✓ | Waste sector interdependency |
| Incineration of waste: Oil/Hazardous waste - Wastewater treatment, other | 0.579 | *x* | Waste sector interdependency |
| Wastewater treatment, food - Landfill of waste: Textiles | 0.565 | ✓ | Waste sector interdependency |
| Manufacture of motor vehicles, trailers and semi-trailers - Production of electricity by geothermal | 0.556 | ✓ | Transport-Energy sectors dependency |
| Other land transport - Air transport | 0.554 | ✓ | Transport sector interdependency |
| Manufacture of motor vehicles, trailers and semi-trailers - Manufacture of other transport equipment | 0.553 | ✓ | Transport sector interdependency |
| Production of electricity by solar thermal - Production of electricity by geothermal | 0.545 | *x* | Energy sector interdependency |
| Wastewater treatment, other - Landfill of waste: Inert / Metal / Hazardous | 0.538 | ✓ | Waste sector interdependency |
| Production of electricity by nuclear - Collection, purification and distribution of water | 0.537 | *x* | Energy-Water sectors dependency |

| Distribution and trade of electricity - Other land transport | 0.492 | ✓ | Transport-Energy sectors dependency |
|---|---|---|---|
| Manufacture of other transport equipment - Transport via pipelines | 0.489 | ✓ | Transport sector interdependency |
| Production of electricity by nuclear - Air transport | 0.488 | x | Transport-Energy sectors dependency |
| Incineration of waste: Food - Incineration of waste: Plastic | 0.487 | x | Waste sector interdependency |
| Incineration of waste: Food - Incineration of waste: Textiles | 0.487 | x | Waste sector interdependency |
| Manufacture of motor vehicles, trailers and semi-trailers - Transport via railways | 0.484 | ✓ | Transport sector interdependency |
| Transmission of electricity - Other land transport | 0.483 | ✓ | Transport-Energy sectors dependency |
| Production of electricity by nuclear - Transmission of electricity | 0.481 | x | Energy sector interdependency |
| Production of electricity by nuclear - Other land transport | 0.477 | x | Transport-Energy sectors dependency |
| Collection, purification and distribution of water - Other land transport | 0.474 | ✓ | Transport-Water sectors dependency |
| Production of electricity by biomass and waste - Transport via pipelines | 0.468 | ✓ | Transport-Energy sectors dependency |
| Manufacture of motor vehicles, trailers and semi-trailers - Production of electricity by wind | 0.464 | ✓ | Transport-Energy sectors dependency |
| Production of electricity by petroleum and other oil derivatives - Other land transport | 0.463 | ✓ | Transport-Energy sectors dependency |
| Production of electricity not elsewhere classified - Wastewater treatment, other | 0.462 | ✓ | Energy-Waste sectors dependency |
| Production of electricity by wind - Collection, purification and distribution of water | 0.462 | x | Energy-Water sectors dependency |
| Production of electricity by petroleum and other oil derivatives - Distribution and trade of electricity | 0.475 | ✓ | Energy sector interdependency |
| Production of electricity not elsewhere classified - Wastewater treatment, food | 0.455 | x | Energy-Waste sectors dependency |
| Steam and hot water supply - Wastewater treatment, other | 0.449 | x | Water-Waste sectors dependency |
| Production of electricity by wind - Transmission of electricity | 0.444 | x | Energy sector interdependency |
| Production of electricity by coal – Production of electricity by petroleum and other oil derivatives | 0.442 | ✓ | Energy sector interdependency |
| Production of electricity by wind - Transport via pipelines | 0.440 | x | Energy sector interdependency |
| Composting of food waste, incl. land application - Biogasification of sewage slugde, incl. land application | 0.434 | x | Waste sector interdependency |
| Sea and coastal water transport - Incineration of waste: Wood | 0.430 | x | Transport-Waste sectors dependency |
| Biogasification of paper, incl. land application - Biogasification of sewage slugde, incl. land application | 0.430 | x | Waste sector interdependency |
| Incineration of waste: Metals and Inert Materials - Wastewater treatment, other | 0.428 | x | Waste sector interdependency |
| Composting of food waste, incl. land application - Wastewater treatment, other | 0.420 | x | Waste sector interdependency |
| Production of electricity by solar photovoltaic- Production of electricity by solar thermal | 0.420 | x | Energy sector interdependency |

| Production of electricity by nuclear - Distribution and trade of electricity | 0.410 | *x* | Energy sector interdependency |
|---|---|---|---|
| Production of electricity by nuclear - Production of electricity by petroleum and other oil derivatives | 0.409 | *x* | Energy sector interdependency |
| Incineration of waste: Oil/Hazardous waste - Wastewater treatment, food | 0.407 | *x* | Waste sector interdependency |
| Sale, maintenance, repair of motor vehicles, motor vehicles parts, motorcycles, motorcycles parts and accessories - Landfill of waste: Paper | 0.407 | *x* | Transport-Waste sectors dependency |
| Transport via railways - Sale, maintenance, repair of motor vehicles, motor vehicles parts, motorcycles, motorcycles parts and accessories | 0.406 | ✓ | Transport sector interdependency |
| Production of electricity by gas - Post and telecommunications | 0.405 | ✓ | Energy-Communication sectors dependency |
| Production of electricity by solar photovoltaic - Transport via railways | 0.401 | *x* | Transport-Energy sectors dependency |
| Sale, maintenance, repair of motor vehicles, motor vehicles parts, motorcycles, motorcycles parts and accessories - Landfill of waste: Food | 0.396 | ✓ | Energy-Waste sectors dependency |
| Sale, maintenance, repair of motor vehicles, motor vehicles parts, motorcycles, motorcycles parts and accessories - Landfill of waste: Textiles | 0.387 | ✓ | Transport-Waste sectors dependency |
| Production of electricity by gas - Transport via railways | 0.385 | ✓ | Transport-Energy sectors dependency |
| Composting of food waste, incl. land application - Composting of paper and wood, incl. land application | 0.381 | *x* | Waste sector interdependency |
| Wastewater treatment, food - Landfill of waste: Wood | 0.381 | *x* | Waste sector interdependency |
| Biogasification of paper, incl. land application - Composting of paper and wood, incl. land application | 0.377 | *x* | Waste sector interdependency |

* ✓ means fewer than 10% of the data points used in the correlation are missing. *x* means more than 10% of the data points used are missing.

** Production of electricity by hydro does not show any correlation. Incineration, biogasification and composting of waste data is not sufficient.

A key finding is that EXIOBASE provides comprehensive data for landfill waste disposal and wastewater treatment, but lacks sufficient coverage across regions for waste incineration, biogasification, and composting. Another key finding is that this type of research can be replicated with a focus on the energy sector, where numerous interdependencies emerge (see Table 1). However, it cannot be effectively extended to the water or communication sectors due to the limited representation of subsectors within EXIOBASE.

**3. Results**

The results of our analysis are summarized in a sectoral interdependency map, where each sector is represented by a distinct colour code (Figure 3): brown for waste, blue for water, green for energy, and red for transport. Sub-sectors are illustrated within large circles corresponding to these colour classifications. The interconnections between sectors are visualized through color-coded lines: brown lines indicate interdependencies within the waste sector, blue lines capture relationships between waste and water, green lines represent dependencies between waste and energy, and red lines highlight the connections between waste and transport.

Sectors with relatively small samples, which produced statistically non-robust results, are marked in magenta. These are not included in Figure 4 of the paper due to their limited robustness, but are retained in Figure 3 for comparative purposes and to illustrate potential future avenues for more refined analysis.

A first inspection of the visualization reveals that all sectors interact, either directly or indirectly. Some relationships are straightforward—for instance, direct ties between waste processing and water use—whereas others are mediated through intermediate processes. An illustrative case is railway emissions, which are not directly connected to other sectors; however, they are indirectly linked through transport manufacturing, itself strongly associated with multiple other sectors. This highlights the importance of examining both direct and indirect systemic linkages in order to capture the broader network of dependencies

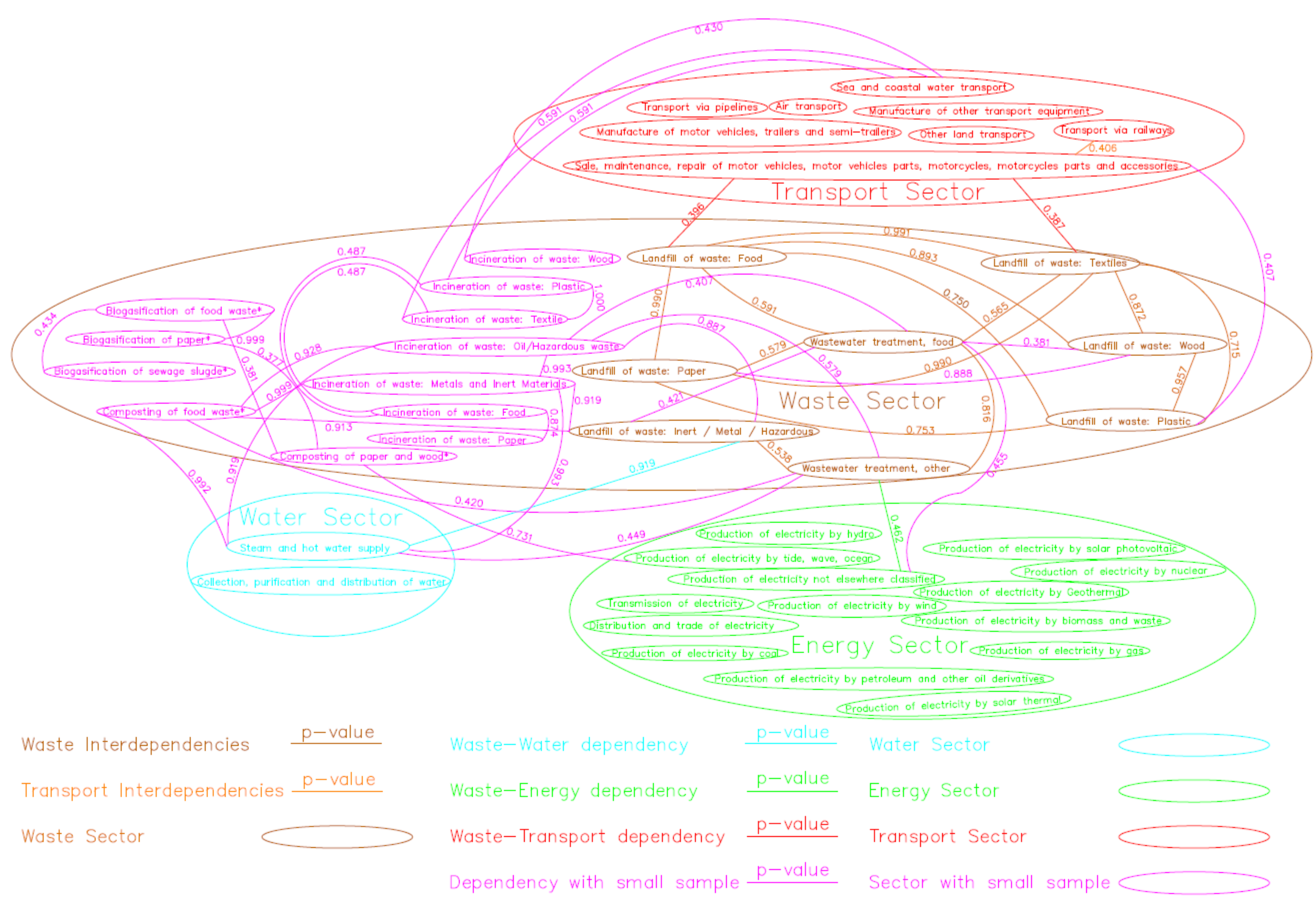

Figure 3 EXIOBASE Emission correlation of waste within and across sectors (all data)

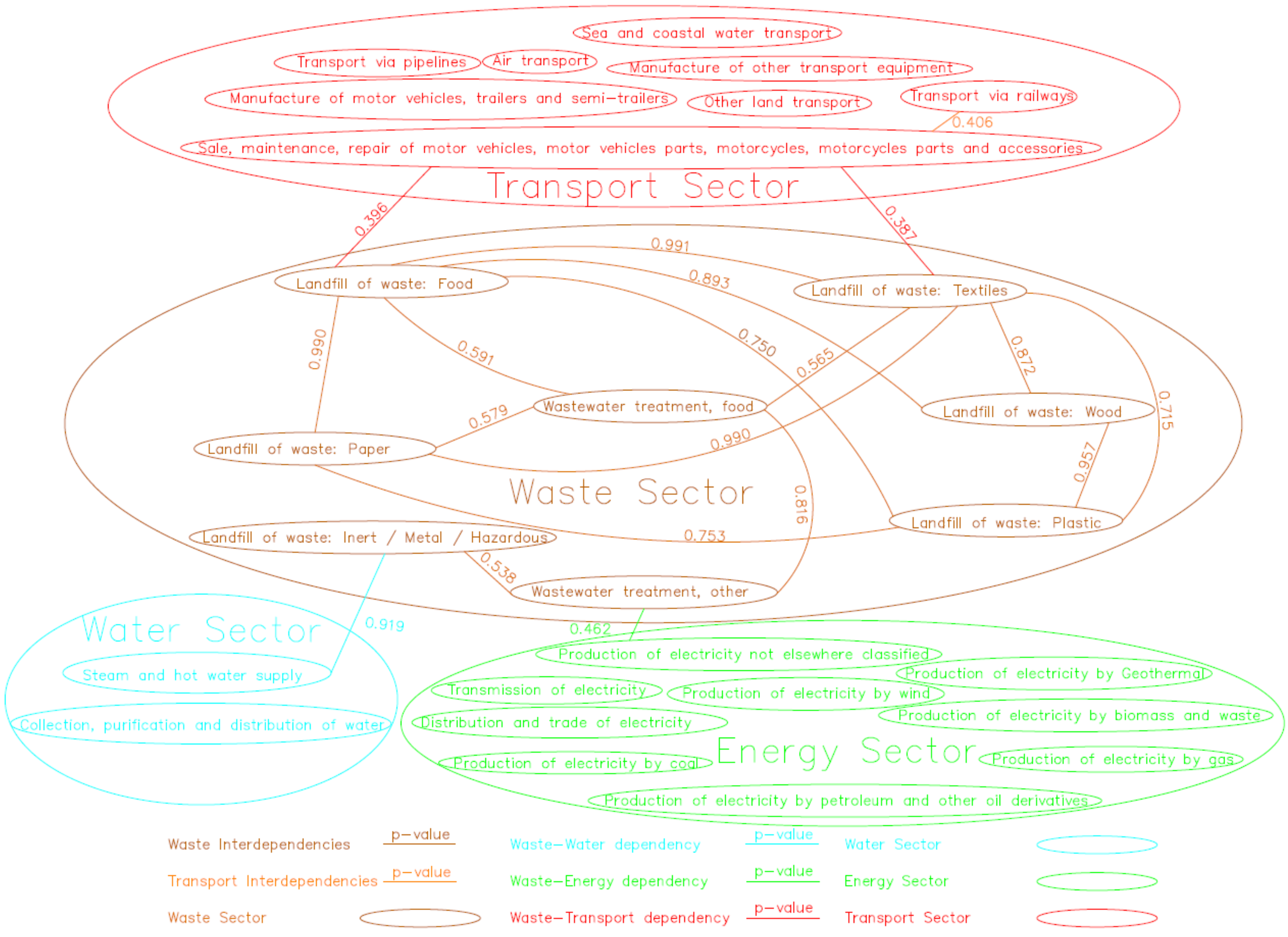

Figure 4 EXIOBASE Emission correlation of waste within and across sectors (robust data)

**4. Theoretical framework and discussion**

Correlations are categorized as high ($r \geq 0.9$), moderate ($0.7 < r < 0.9$), and weak ($r \leq 0.7$). Data robustness is indicated by <10% missing data points or >10%, affecting reliability. A more detailed examination of the waste sector, based on correlation analysis, reveals differentiated levels of overlap in emission drivers:

- Landfill waste from food vs. textiles (p-value = 0.991) and paper (p-value = 0.990):

EXIOBASE 3 links emissions to specific material flows (e.g., food, textiles, paper). These high correlations suggest that landfill emissions from food, textiles, and paper share similar drivers, such as organic content leading to methane ($CH_4$) and $CO_2$ emissions during decomposition. EXIOBASE 3's emission accounts for landfill activities (e.g., $CO_2$ equivalents from waste sectors) can capture these overlaps, as food, textiles, and paper often have comparable organic fractions and degradation patterns. Organic waste (e.g., food, textiles, paper) in landfills generates significant methane emissions due to anaerobic decomposition (Ragab et al., 2025; Madhavaraj & Karthikeyan, 2025; Diamond, 2024; Riman et al., 2022; Aziz et al., 2020; Dos Santos et al., 2020; Abouri et al., 2016). The high correlation aligns with shared biochemical processes, as these materials have similar carbon content and decomposition rates. EXIOBASE 3's data, which includes greenhouse gas emissions by waste type, supports this, as landfill emissions for these categories are often modelled with similar emission factors.

Diamond (2024) emphasizes that organic waste—such as food scraps, textiles, and paper—is a major driver of methane emissions in landfills due to anaerobic decomposition (Diamond, 2024). When buried without oxygen, these biodegradable materials undergo microbial processes that release methane ($CH_4$) and carbon dioxide ($CO_2$) (Aziz et al., 2020). Methane is especially concerning, as it is 28–36 times more potent than $CO_2$ in trapping heat over a 100-year timescale (Diamond, 2024). Riman et al. (2022) provide quantitative evidence of how seasonal conditions amplify these processes (Riman et al., 2022). Food waste, paper/textiles, garden waste, and wood were identified as the dominant organic fractions (Diamond, 2024; Riman et al., 2022). The researchers found that wet season conditions enhanced decomposition, resulting in 0.331 Gg of methane compared to 0.134 Gg in the dry season—a 42.2% increase (Riman et al., 2022). This is explained by higher microbial activity under wetter conditions, which accelerates the conversion of degradable organic carbon into methane. Studies that while food and garden waste degrade rapidly, paper and textiles decompose at moderate rates, yet all contribute to the shared pool of degradable organic carbon that sustains methane generation (Ragab et al., 2025; Madhavaraj & Karthikeyan, 2025; Diamond, 2024; Riman et al., 2022; Aziz et al., 2020). Since organic waste is a dominant fraction of municipal solid waste, it plays a central role in making landfills one of the largest anthropogenic methane sources worldwide.

Food, textiles, and paper share similar biochemical pathways of decay, beginning with the hydrolysis of complex carbohydrates, proteins, and cellulose into simpler compounds. These compounds are then fermented into volatile fatty acids, hydrogen, and $CO_2$, which methanogenic bacteria convert into methane. Textiles of natural origin (e.g., cotton, wool) and paper products behave much like food waste because of their organic, cellulose-based structure. This commonality means that mixed organic fractions in landfills collectively accelerate methane generation when oxygen is depleted.

Some studies emphasize that unmanaged landfills in developing regions intensify the climate burden due to lack of methane recovery systems (Diamond, 2024; Riman et al., 2022). Riman et al. (2022) illustrate that methane emissions are effectively equal to methane generated, as open dumpsites lack capture or oxidation processes. Similarly, Diamond (2024) stresses that globally, landfills are among the top methane sources, with outputs depending on organic waste composition, moisture, and landfill design (Diamond, 2024). For example, wetter conditions favour faster microbial activity, leading to higher methane output, while poorly managed sites without gas capture exacerbate climate impacts (Diamond, 2024). On average, a ton of municipal solid waste in a landfill can generate between 0.5 and 2.5 cubic meters of methane annually, underscoring the cumulative effect of discarded organic material (Diamond, 2024). This convergence highlights that organic fractions not only share decay pathways but also dominate overall emission patterns. To mitigate these impacts, Diamond (2024) recommends alternatives such as composting, anaerobic digestion, and waste-to-energy technologies. Composting, for instance, allows aerobic decomposition of food, textiles, and paper, thereby preventing methane release and producing useful soil amendments (Diamond, 2024). Anaerobic digestion, by contrast, controls the same biochemical processes that occur in landfills but captures the methane as biogas for energy use. Thus, addressing the shared decomposition pathways of organic waste is central to reducing the climate burden of landfills.

- Landfill waste from wood and plastic (p-value = 0.957):

Wood and plastic waste in landfills may correlate due to shared disposal pathways or material properties (e.g., both are carbon-based, though plastic is less biodegradable). The high correlation could reflect common waste management practices or co-disposal in similar landfill types, captured in EXIOBASE 3's waste treatment data. Wood and certain plastics (e.g., non-biodegradable polymers) in landfills contribute to long-term carbon storage or slow-release $CO_2$, with emissions influenced by landfill conditions (e.g., moisture, compaction) (Ingle et al., 2025; Bugarčić et al., 2024; Singh & Chunglok, 2022; Meereboer et al., 2020; Souza et al., 2020; Bohlmann, 2005). The correlation may stem from these shared environmental factors, which EXIOBASE 3's emission accounts can quantify.

Wood and other lignocellulosic residues disposed of in landfills exhibit slow degradation due to their lignin-rich, recalcitrant structure (Ingle et al., 2025). Wood, composed mainly of cellulose and lignin, biodegrades but very slowly in the oxygen- and moisture-limited landfill environment (Bugarčić et al., 2024). While anaerobic conditions can convert part of this carbon into methane and carbon dioxide, a significant fraction remains stored over the long term, effectively functioning as a carbon sink (Ingle et al., 2025; Singh & Chunglok, 2022). Conventional plastics, in contrast, resist degradation and persist for centuries, while bioplastics such as PLA, PBS, and PHAs—derived from renewable feedstocks—are designed to degrade, though their breakdown remains limited in landfills (Bugarčić et al., 2024; Meereboer et al., 2020). Their main environmental risk lies not in gaseous emissions but in fragmentation into microplastics (Ingle et al., 2025). Wood and plastic wastes, though differing in biodegradability, share carbon-based composition and often enter landfills through similar disposal pathways. The extent of emissions from wood-like biomass is strongly influenced by landfill conditions: higher moisture levels accelerate microbial activity and gas generation, whereas dryness and compaction favour carbon storage (Ingle et al., 2025). Plastics, however, remain largely inert irrespective of such conditions. Both therefore function as long-term carbon storage pools, with eventual $CO_2$ release—and under anaerobic conditions, methane—depending on landfill conditions like compaction, microbial activity, and moisture (Bugarčić et al., 2024; Meereboer et al., 2020). Bioplastics bridge the gap—renewable like wood, but functionally plastics—though their biodegradability is limited in typical landfill environments.

- Landfill waste from food vs. wood (p-value = 0.893) and plastic (p-value = 0.750); textiles vs. wood (p-value = 0.872) and plastic (p-value = 0.715):

Moderate correlations suggest partial overlaps in emission drivers. Food and wood share organic properties, but wood decomposes more slowly, leading to less pronounced correlations. EXIOBASE 3 tracks these materials through industries like "Food products" and "Textiles," with emissions linked to landfill processes, supporting these moderate relationships. Food and wood have different methane yields in landfills (Merab et al., 2025; Bhuiya et al., 2025; Ragab et al., 2025; Santos et al., 2024; Sohoo et al., 2021; Tesseme & Chakma, 2021; El-Fadel & Massoud, 2000; Owens & Chynoweth, 1993), explaining the moderate correlation. Plastic's lower biodegradability reduces its emission correlation with organic wastes (Pekrioğlu Balkıs & Ahmad, 2025; Ragab et al., 2025; Moshood et al., 2022; Sikorska et al., 2021; Ferrand et al., 2020). EXIOBASE 3's data, which differentiates emission intensities by material, aligns with these findings.

Food and wood wastes behave differently in landfills due to their distinct biochemical properties, which in turn influences their methane generation potential. Food waste makes up a significant portion of the municipal solid waste stream (about 32%) and contains a high fraction of degradable organic carbon (Merab et al., 2025). Because it decomposes quickly under anaerobic conditions, food waste generates methane relatively rapidly and in large quantities (Merab et al., 2025). This makes it one of the main contributors to landfill methane emissions (Merab et al., 2025).

Wood waste, by contrast, is present in very small amounts (only about 0.1% of the waste stream) and is highly resistant to decay because of its lignin-rich structure (Merab et al., 2025). In methane generation models, wood is assigned a much lower degradable organic carbon coefficient (0.03 compared to 0.15 for food), reflecting its slow decomposition and minimal methane yield (Merab et al., 2025). As a result, wood contributes negligibly to landfill methane emissions, both because of its small share in the waste composition and because it breaks down at a much slower rate.

Food waste, rich in easily degradable carbohydrates, fats, and proteins, breaks down rapidly under anaerobic conditions, leading to high methane yields (Wang et al., 2018). In contrast, wood waste is primarily composed of lignocellulosic material, which resists microbial degradation (Tongbuekeaw et al., 2021). As a result, wood generates methane more slowly and in smaller quantities (Tongbuekeaw et al., 2021). These differences explain why correlations between food and wood methane emissions are only moderate rather than strong, as the rate and magnitude of gas production vary significantly between the two waste streams.

Plastic waste, on the other hand, exhibits a much lower degree of biodegradability compared to food and wood, and this substantially alters its relationship to landfill methane emissions (Bugarčić et al., 2024; Meereboer et al., 2020). Most plastics are chemically stable and resist microbial breakdown, meaning their direct contribution to methane production is minimal (Bugarčić et al., 2024; Meereboer et al., 2020). While certain biodegradable plastics can degrade under specific conditions, their impact remains relatively small compared to organic materials (Meereboer et al., 2020). This low biodegradability leads to weaker correlations between plastic disposal and methane emissions when compared to organic wastes, reinforcing the idea that plastic’s role in landfill gas generation is secondary rather than primary.

- Wastewater treatment for food vs. other waste types (p-value = 0.816) and landfill of inert/metal/hazardous waste (p-value = 0.816):

The moderate correlation of wastewater treatment for food with other waste types suggests shared treatment processes (e.g., anaerobic digestion) or co-treatment of effluents. The correlation with inert/metal/hazardous waste landfills may reflect integrated waste management systems where wastewater and landfill emissions are linked via regional infrastructure, captured in EXIOBASE 3's regional data. Wastewater treatment emissions notes that food processing effluents often undergo similar treatment processes (e.g., aerobic/anaerobic digestion) as other organic wastes (Talaiekhozani, 2019; Zheng et al., 2013; Kosseva, 2009), leading to moderate correlations. The link to inert/hazardous waste may reflect co-treatment in facilities handling mixed waste streams.

Food industry wastewater, including effluents from dairy, meat, fermentation, and seafood processing, is highly biodegradable and comparable to other organic-rich waste streams (Talaiekhozani, 2019; Kosseva, 2009). These effluents typically contain proteins, fats, carbohydrates, and cleaning chemicals, leading to high biochemical oxygen demand and chemical oxygen demand. Consequently, biological processes such as anaerobic digestion, aerobic treatment, and combined anaerobic–aerobic systems are widely applied. Similar processes are used in treating other organic wastes like agricultural residues and sludge, enabling co-treatment strategies. Anaerobic co-digestion of food effluents with additional substrates has been shown to improve methane yield, nutrient balance, and process stability (Talaiekhozani, 2019; Kosseva, 2009).

The overlap between wastewater treatment and landfill management reflects the push toward integrated waste systems. Food waste that ends up in landfills releases methane, while effluents from food processing and landfill leachates share high organic loads and trace contaminants (Kosseva, 2009). This creates opportunities for shared infrastructure, where anaerobic digesters or hybrid biological systems can treat both wastewaters and leachates. European Union policies, such as the Landfill Directive, have further driven this integration by restricting biodegradable waste disposal in landfills and encouraging valorisation pathways like anaerobic digestion, composting, or conversion into added-value products (Kosseva, 2009).

At the emissions level, both food effluents and landfill leachates contribute organic pollution, nutrients, and sometimes metals, requiring similar treatment strategies. Despite challenges such as heavy metals or detergents that can inhibit biological processes, optimization methods have enabled stable operation of digesters (Kosseva, 2009). Crucially, anaerobic systems not only reduce pollution but also recover energy in the form of methane and facilitate nutrient recycling (Talaiekhozani, 2019; Kosseva, 2009). This aligns food wastewater management with broader circular economy principles, where waste is viewed as a feedstock for energy and resource recovery, linking food processing, wastewater, and solid waste streams within integrated regional infrastructures.

- Landfill waste from food/textiles/paper vs. wastewater treatment for food (p-values = 0.591, 0.565, 0.579); paper vs. plastic (p-value = 0.753):

Landfill emissions (primarily methane and $CO_2$ from decomposition) differ significantly from wastewater treatment emissions (e.g., $CO_2$ from aeration or $N_2O$ from denitrification). EXIOBASE 3 separates these processes under different sectors (e.g., "Waste treatment" vs. "Sewage disposal"), explaining the low correlation. Similarly, paper and plastic have different landfill behaviours (paper is biodegradable, plastic is not), reducing their emission correlation. Landfill and wastewater treatment emissions have distinct mechanisms (e.g., methane from landfills vs. nitrous oxide from wastewater) (Chidambarampadmavathy et al., 2017; Ashrafi et al., 2015). Paper and plastic's correlation aligns with their differing biodegradability, as noted in waste management studies (Markevičiūtė, 2025; Gómez & Michel 2013; Guillet et al., 1992). EXIOBASE 3's emission accounts reflect these differences through sector-specific data.

When comparing emissions from landfills and wastewater treatment, both papers emphasize that the mechanisms and gas profiles differ significantly. Landfills are dominated by methane ($CH_4$) and carbon dioxide ($CO_2$) generated through anaerobic decomposition of biodegradable organic waste, with paper contributing heavily due to its degradability, while plastics remain inert and do not emit significant greenhouse gases (GHGs) during landfill disposal (Chidambarampadmavathy et al., 2017; Ashrafi et al., 2015). Wastewater treatment, on the other hand, produces $CO_2$ through aeration in aerobic processes and releases nitrous oxide ($N_2O$) during nitrification–denitrification, with anaerobic units contributing additional $CH_4$ emissions. Thus, the emissions are not only different in magnitude but also in composition, with landfills being major methane sources and wastewater systems generating a wider range of gases including $N_2O$, which has a much higher global warming potential (Ashrafi et al., 2015). Chidambarampadmavathy et al. (2017) highlight that plastics, though not biodegradable in landfills, create indirect emission challenges by limiting the degradation of organic matter and contributing to toxic leachates (Chidambarampadmavathy et al., 2017). Importantly, it introduces landfill methane recycling as a mitigation strategy, where $CH_4$ emissions from landfills can be captured and used by methanotrophs to produce polyhydroxyalkanoates (PHAs), a class of biodegradable plastics (Chidambarampadmavathy et al., 2017). his approach not only reduces uncontrolled methane emissions but also offers a cradle-to-cradle solution by transforming landfill gases into sustainable bio-based materials. Together, these studies show that landfill and wastewater emissions operate through distinct pathways: landfills through organic decay yielding $CH_4$ and $CO_2$, and wastewater through biological treatment pathways yielding $CO_2$ and $N_2O$. Moreover, while paper waste amplifies landfill methane emissions due to biodegradability, plastics do not directly contribute but can instead be managed through innovative methane-to-bioplastics technologies (Chidambarampadmavathy et al., 2017; Ashrafi et al., 2015).

To conclude:

- High correlations (p-value > 0.9): Strong overlaps in emission drivers are observed between landfill waste from food, textiles, and paper, as well as between landfill waste from wood and plastic. These high values suggest that these waste types are strongly influenced by common underlying processes.
- Moderate correlations (0.7 ≤ p-value ≤ 0.9): Partial overlaps in emission drivers are found across several combinations, including landfill waste from food, wood, and plastic; landfill waste from textiles, wood, and plastic; wastewater treatment for food and other waste categories; and landfill disposal of inert, metal, and hazardous waste. These correlations indicate a degree of shared drivers, but with sector-specific differences remaining important.
- Weak correlations (p-value < 0.7): Distinct emission processes are observed in cases such as landfill waste from food, textiles, and paper compared with wastewater treatment for food, as well as between landfill waste from paper and plastic. These weak correlations underline the heterogeneity of emission drivers across waste categories.

Taken together, the results demonstrate a complex web of interdependencies both within the waste sector and across sectors (see Figure 3 and Figure 4). The degree of correlation varies significantly depending on the material stream, underscoring the need for tailored mitigation strategies that account for both shared and distinct emission pathways.

### 4.1 Exploring the boundaries of linear modelling in EXIOBASE

An attempt was then made to model the results using linear regression analysis. However, diagnostic testing revealed significant multicollinearity, as indicated by a Variance Inflation Factor (VIF) exceeding 10. To further address the issue of multicollinearity, Principal Component Analysis (PCA) was applied. PCA enables the transformation of a set of highly correlated variables into a smaller set of uncorrelated "virtual variables" (principal components). This procedure helps to reduce redundancy and facilitates the interpretation of relationships within the dataset. PCA's Component Matrix and Rotated Component Matrix (Rotation Method: Varimax with Kaiser Normalization) illustrate the degree to which each original variable contributes to the newly generated principal components (Field, 2024). An important property of PCA is that, when the units of the original variables are consistent (e.g., monetary values, emissions, or physical measurements), the derived components may be interpreted as retaining the same unit, thereby preserving comparability. Despite these advantages, the application of PCA in this study revealed certain challenges. Specifically, the dimensionality of the data remained high even after transformation, resulting in a relatively large number of principal components. Several factors may explain this outcome such as high dimensionality:

- EXIOBASE includes a large number of original variables, many of which exhibit complex interdependencies, nonlinear relationships
- PCA is inherently a linear technique and may fail to capture nonlinear patterns present in the data and presence of noise
- Variability and uncertainty in the dataset may reduce the ability of PCA to effectively condense information.

This outcome demonstrated the limitations of linear regression in this context and underscored the necessity of alternative methodological approaches. Consequently, the research plan focuses on developing models grounded in dependency analysis, which offer greater capacity to capture the complex structure of the data and provide a systematic understanding of sectoral interactions and emission-reduction potentials.

**5. Conclusions**

The analysis presented in this study highlights the systemic interdependencies of emissions across waste management, energy, transport, water, and communication sectors. By leveraging the EXIOBASE 3 database, the research has identified clear patterns of correlation within and across waste categories, with organic waste streams such as food, textiles, and paper exhibiting particularly high overlaps in emission drivers. Moderate and weak correlations across other categories reveal the differentiated nature of waste decomposition and treatment processes, underscoring the complexity of emissions pathways. These findings illustrate that while certain mitigation strategies can target common drivers, sector- and material-specific approaches remain necessary to address heterogeneity.

From a methodological perspective, the comparative evaluation of process-based LCA and EXIOBASE underscores the complementarity of product-level granularity and system-wide coverage. While process-based approaches capture detailed product-level processes, EXIOBASE enables the integration of emissions across global supply chains, thus supporting systemic analysis of intersectoral dependencies. Attempts to apply linear regression and PCA to the EXIOBASE dataset revealed the limitations of conventional linear models in capturing the nonlinear and multidimensional character of interdependencies. This reinforces the importance of advancing methodological frameworks that are capable of accommodating complexity, particularly for policy-relevant applications in climate mitigation and resource management.

Overall, the study demonstrates that emissions from waste management cannot be understood in isolation but are embedded within broader economic and infrastructural systems. High levels of interconnection between waste, energy, and water systems point to the potential for integrated mitigation strategies that maximize co-benefits across sectors. At the same time, the persistence of sector-specific patterns suggests that policies must balance systemic coordination with targeted interventions. The research thereby contributes to both academic understanding and policy design by providing empirical evidence of emission linkages and highlighting methodological pathways for more robust analysis of sectoral interdependencies.

**Acknowledgements**

The authors gratefully acknowledge the University of Thessaly, the McGill University, the Université Laval and the financial support of the UK Engineering and Physical Sciences Research Council under grant number EP/R017727 (UK Collaboratorium for Research on Infrastructure and Cities).